\documentclass[a4paper]{spie}

\usepackage[sc, hang]{caption}
\usepackage{graphicx}
\usepackage{graphics}
\usepackage{makeidx}
\usepackage[pass]{geometry}
\usepackage{booktabs}
\usepackage{marvosym}
\usepackage[hang]{subfigure}
\usepackage{SIunits}
\usepackage{rotating}
\usepackage{csquotes}
\usepackage{url}
\usepackage{multicol}
\usepackage{multirow}
\usepackage{overpic}
\usepackage{amsmath}
\usepackage{commath}
\usepackage{bm}
\usepackage{rotating}
\usepackage{psfrag}
\usepackage{parallel}
\usepackage{fancyhdr}
\usepackage[breaklinks]{hyperref}

\usepackage[english]{babel} 
\usepackage{eso-pic,graphicx}
\usepackage{enumerate}
\usepackage{psfrag}
\usepackage{booktabs}
\usepackage{hyperref}      
\hypersetup{               
   pdftitle={mavisSPIE2022},    
   pdfauthor={GuidoAgapito},  
   colorlinks=true,      
   linkcolor=black,      
   anchorcolor=black,    
   citecolor=black,      
   filecolor=black,      
   menucolor=black,      
   pagecolor=black,      
   urlcolor=blue,         
   linktocpage=true
}

\usepackage{color, colortbl}
\definecolor{LightGray}{gray}{0.9}
\usepackage{diagbox}
\usepackage{array}
\newcolumntype{C}[1]{>{\centering\let\newline\\\arraybackslash\hspace{0pt}}m{#1}}

\title{MAORY/MORFEO and rolling shutter induced aberrations in laser guide star wavefront sensing}

\author[a,c]{Guido Agapito}
\author[a,c]{Lorenzo Busoni}
\author[a,c]{Giulia Carlà}
\author[a,c]{Cédric Plantet}
\author[a,c]{Simone Esposito}
\author[b]{Paolo Ciliegi}
\affil[a]{INAF -- Osservatorio Astrofisico di Arcetri, Largo E. Fermi 5, 50125, Firenze, Italy}
\affil[b]{INAF -- Osservatorio di Astrofisica e Scienza dello Spazio di Bologna (OAS), via Gobetti 93/3, Bologna, Italy}
\affil[c]{ADaptive Optics National laboratory in Italy (ADONI)}

\authorinfo{Further author information:\\Guido Agapito, \Letter
 \hspace{0.5ex} guido.agapito@inaf.it\\}

\begin{document} 
\maketitle

\begin{abstract}
Laser Guide Star (LGS) Shack-Hartmann (SH) wavefront sensors for next generation Extremely Large Telescopes (ELTs) require low-noise, large format ($\sim$1Mpx), fast detectors to match the need for a large number of subapertures and a good sampling of the very elongated spots. One path envisaged to fulfill this need has been the adoption of CMOS detectors with a rolling shutter read-out scheme, that allows low read-out noise and fast readout time at the cost of image distortion due to the detector rows exposed in different moments. In this work we analyze the impact of the rolling shutter read-out scheme when used for LGS SH wavefront sensing of the Multiconjugate adaptive Optic Relay For ELT Observations (MORFEO, formerly known as MAORY) for ESO ELT; in particular, we focus on the impact on the adaptive optics correction of the distortion-induced aberrations created by the rolling exposure in the case of fast varying aberrations, like the ones coming from the LGS tilt jitter due to the up-link propagation of laser beams. We show that the LGS jitter-induced aberration for MORFEO can be as large as 100nm rms and we discuss possible mitigation strategies.
\end{abstract}


\section{Introduction}
\label{sect:intro}

Next-generation Extremely Large Telescopes (ELTs)\cite{2020SPIE11445E..08B,2012SPIE.8447E..1JE,2020SPIE11445E..1ET} foresee the use of Laser Guide Stars (LGSs) in their adaptive optics (AO) systems \cite{2020SPIE11448E..0YC,10.1117/12.2231681,2018SPIE10703E..3VC,10.1117/12.2232411,10.1117/12.923486,2013aoel.confE...4C,10.1117/12.2314255}. 
They all use Shack-Hartmann (SH) LGS wavefront sensors (WFSs) that are highly demanding in terms of detector's specifications\cite{Hippler2019,2016SPIE.9909E..5ZG,oberti:hal-02614170,2021A&A...649A.158B}.
In this work we focused on the Multiconjugate adaptive Optic Relay For ELT Observations\cite{2020SPIE11448E..0YC,2021Msngr.182...13C,ciliegi2022} (MORFEO, formerly known as MAORY) and on the impact of rolling shutter detectors in the LGS WFS.
In fact, during phase B of MORFEO development the two options for the LGS WFS detector that satisfy the requirements are: ESO/e2v LISA camera\cite{10.1117/12.2314489,BeleticRadialCCD, DowningNGSD2014, MorenoNGSD2016, FeautrierSPIE2012} with rolling shutter read-out and cameras based on SONY IMX 425 chip with global shutter read-out.
ESO and the MORFEO consortium decided to make a comparison study to evaluate which is the best solution for MORFEO.
As we reported in Agapito, Busoni \textit{et al.} 2022\cite{2022JATIS...8b1505A}, complementary metal oxide semiconductor (CMOS) sensors with a rolling shutter readout are an alternative solution to charge coupled device (CCD) with global shutter, but WFSs provided with such devices are characterized by distortion induced aberration (DIA).
In this paper we present the performance estimation for MORFEO with rolling shutter and global shutter read-out detectors, in terms of the expected SR in the best atmospheric conditions and the sky coverage at the south galactic pole.

In Sec.~\ref{sect:jitter} we quantify the residual LGS tilt in both amplitude and temporal evolution for a median atmospheric condition\cite{2013aoel.confE..89S}, taking into account the MORFEO jitter compensation;
in Sec.~\ref{sec:mitigation} we present the level of distortion induced aberrations and discuss possible mitigation of them;
in Sec.~\ref{sec:prop} we study the propagation of distortion induced aberrations in the MORFEO tomographic reconstructor;
in Sec.~\ref{sec:performance} we present the performance estimation;
in Sec.~\ref{sec:conclusion} we report our conclusion.

\section{LGS jitter}
\label{sect:jitter}
\begin{figure}[h]
    \centering
    \includegraphics[width=0.5\linewidth]{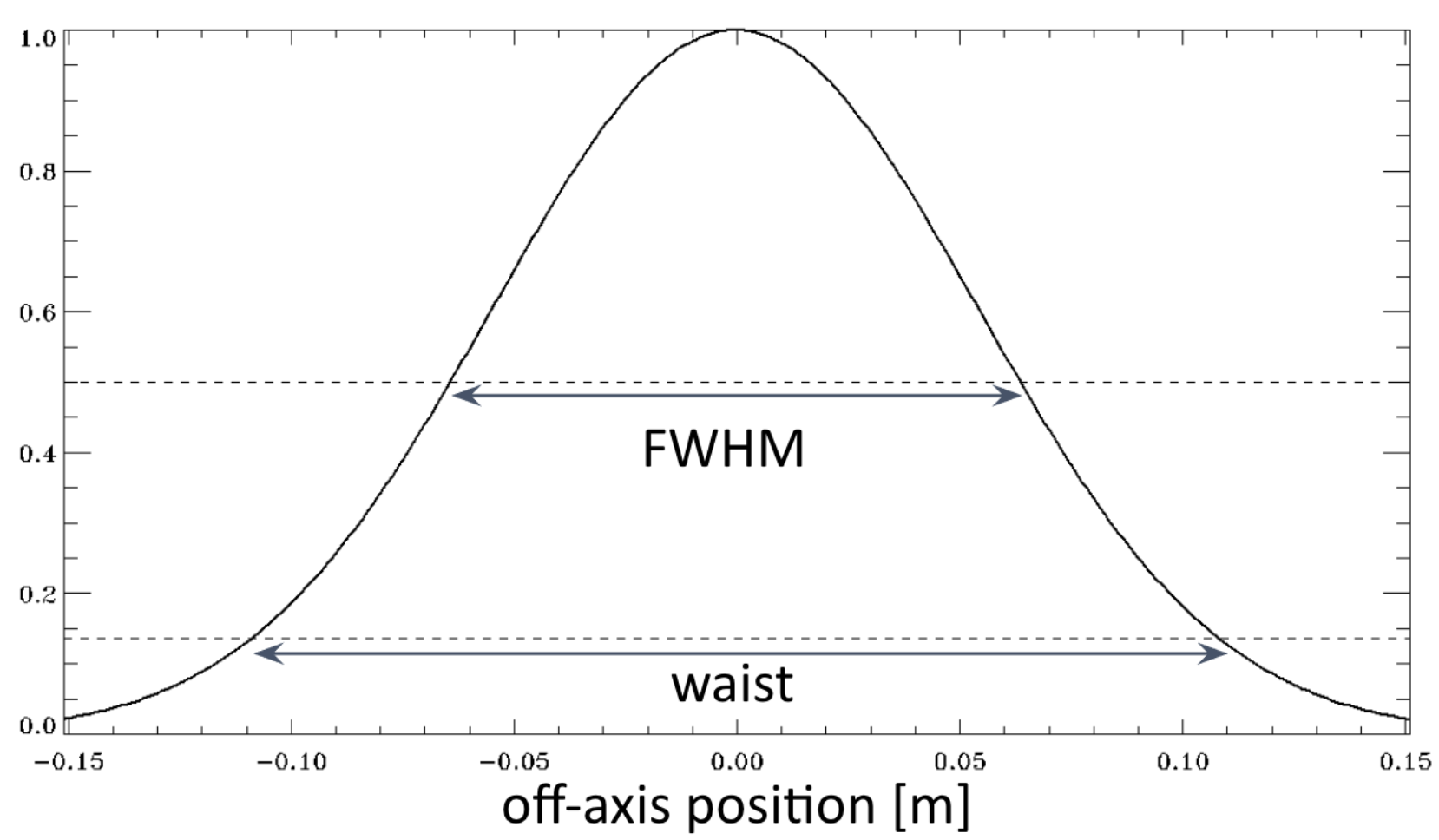}
    \caption{Section of the uplink projection intensity.}
    \label{fig:laser_beam}
\end{figure}

\begin{figure}[h]
    \centering
    \includegraphics[width=0.6\linewidth]{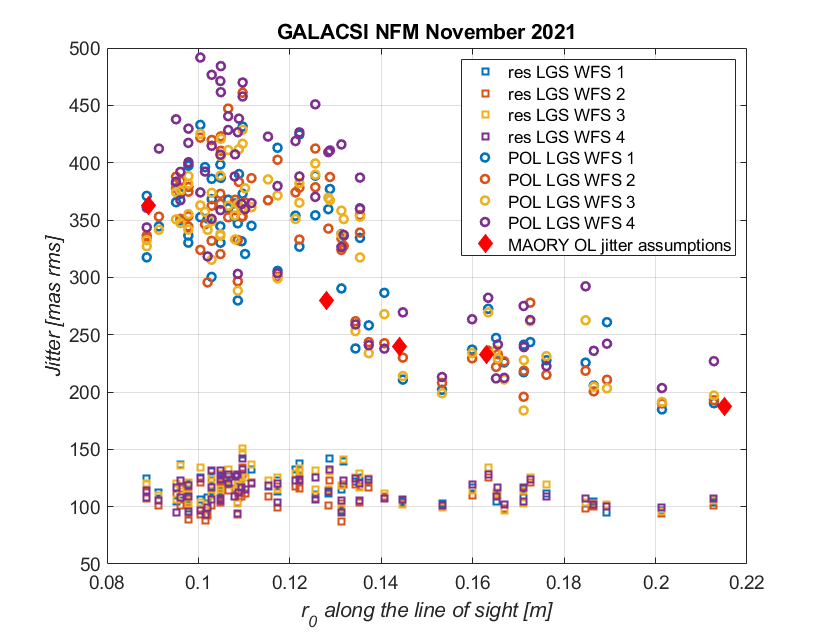}
    \caption{Comparison between estimated open-loop single axis jitter from simulations (red diamonds) and ESO VLT GALACSI data\cite{GALACSI_SYLVAIN}.}
    \label{fig:GALACSI_Jitter}
\end{figure}

\begin{figure}[h]
    \centering
    \includegraphics[width=0.75\linewidth]{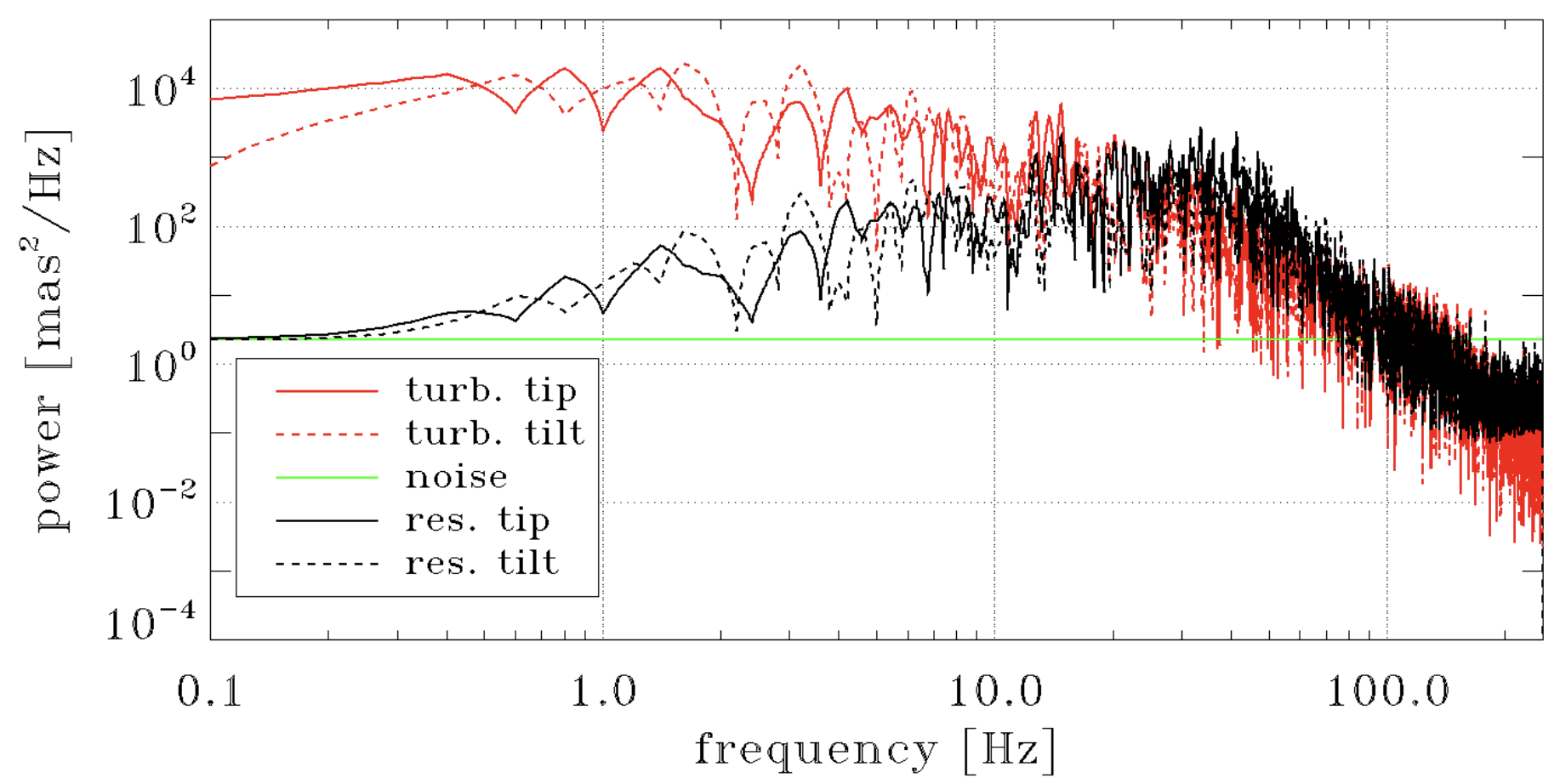}
    \caption{Tip and Tilt PSDs for the upward propagation in open and closed loop.}
    \label{fig:laser_jitter_PSD}
\end{figure}

LGS tilt jitter is determined by turbulence-induced fluctuations introduced during both the upward and downward propagation of the laser beam.
We consider negligible the downward propagation for a 39m telescope, because the upward propagation is characterized by a significantly smaller beam.
We run a set of end-to-end simulations with PASSATA\cite{doi:10.1117/12.2233963} to characterize this jitter and its residual after closed loop correction.
The uplink projection is considered to be a gaussian beam with a waist radius of 108mm and a full aperture of 302mm diameter\cite{2011aoel.confE..56H,2015aoel.confE..45H,2021A&A...649A.158B} (see Fig.\ref{fig:laser_beam}).
We compute the laser beacon image at 90km and we extract with a centroid the open loop jitter: we got 340mas RMS (square sum of the two axes) for the median atmospheric profile\cite{2013aoel.confE..89S}. This value is in good agreement with ESO VLT GALACSI data as can be seen in Fig. \ref{fig:GALACSI_Jitter}.
Then, we compute the closed loop residual considering as input the centroid signal, as control a pure integrator and as plant a pure delay of 3.35 frames (1 due to read-out, 1 due to sample$\&$hold, 0.5 due to real time computer and 0.5 due to deformable mirror response time and 0.35 are the delay due to the round-trip propagation of the laser beam to the sodium layer\footnote{In MORFEO the correction is done using a fast tip/tilt mirror in the launch telescopes.}), with a frame period of 2ms.
We found a residual jitter of 190mas RMS.
The Power Spectral Densities of these signals are reported in Fig.\ref{fig:laser_jitter_PSD}.
In the following sections we will use these results, though it is worth noting that the amount of residual jitter varies according to the atmospheric conditions. 

\section{Mitigation of distortion induced aberrations}\label{sec:mitigation}

We computed the DIA from the residual jitter as described in Ref. \citeonline{2022JATIS...8b1505A} for a detector with rolling shutter read-out on two sectors and we got, for the median profile\cite{2013aoel.confE..89S}, a wavefront error due to DIA of 370 nm rms.
This is an extremely large value and it will compromise the functionality of MORFEO, for this reason we study a mitigation strategy.
A possible mitigation could be based on the forecast of the tilt temporal evolution using the LGS WFS measurements from the current and previous steps.
In fact, knowing accurately the tilt temporal evolution during the current (\emph{i.e.} last) exposure of the frame will allow correction of the DIA.
We analyze in this section this approach and its efficiency.

We considered a linear extrapolation of the speed based on the estimation of the past values of speed and acceleration\cite{LINEST_GAGO}:
\begin{equation}\label{eq:v0}
    v_0[k] = p[k] - p[k-1]
\end{equation}
\begin{equation}\label{eq:a0}
    a_0[k] = v_0[k] - v_0[k-1]
\end{equation}
\begin{equation}\label{eq:v}
    v[k] = v[k] + 0.5 a_0[k]
\end{equation}
\begin{equation}\label{eq:a}
    a[k] = v[k] - v[k-1]
\end{equation}
where $p[k]$ is the average tilt value, $v_0[k]$ is the initial estimation of average tilt speed in nm/frame,  $a_0[k]$ is the initial estimation of average tilt acceleration in nm$^2$/frame, $v[k]$ is the final estimation of average tilt speed in nm/frame, and $a[k]$ is the final estimation of average tilt acceleration in nm$^2$/frame. All these average values refer to the detector integration that ends at the time step $k$.\\
This estimation is used to compute the tilt speed during the detector integration\footnote{please note that it means a forecast of half frame, because the speed is estimated from the average measurements of tilt during the frame integration.} and to remove from the slopes the DIA.
Then we can estimate the error on this mitigation computing the error on the tilt speed: in fact the error of this mitigation strategy is proportional to the error on tilt speed estimation. Please note that the tilt speed is the main contributor to the DIA as shown in Ref. \citeonline{2022JATIS...8b1505A} and we can neglect the DIA due to higher orders.

\begin{figure}[h]
    \centering
    \includegraphics[width=0.55\linewidth]{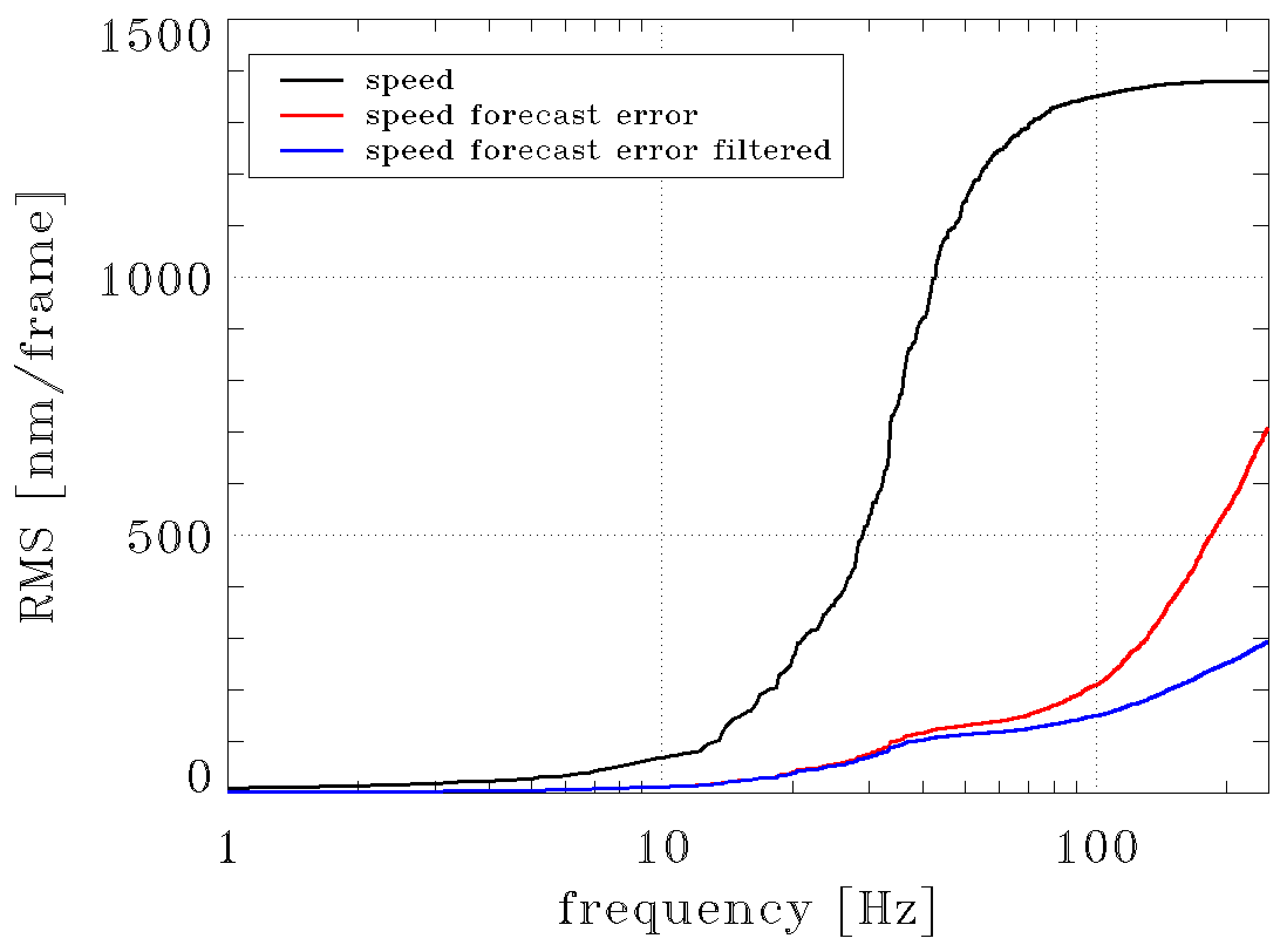}
    \caption{Cumulative RMS of the jitter speed. The black line is the speed signal computed from the phase, the red line is the error on speed considering the forecast done considering the current and previous steps coming for the tip and tilt retrieved from the slopes of a 60$\times$60 SHS (with 15arcsec FoV, elongated spots, windowed CoG, photon noise and read-out noise) and the blue line is the previous error after being filtered by the closed loop..}
    \label{fig:cumulative_RMS_jitter_speed}
\end{figure}

We considered a baseline-like case with 60$\times$60sub-apertures, 15arcsec FoV, elongated spots (``multi-peak'' sodium profile \cite{2014A&A...565A.102P} at z=30deg), windowed CoG, photon noise and read-out noise (considering ~1000ph/frame/sa flux and 3e- RON) to have a more realistic condition even if some details are missing like turbulence and the tomographic closed loop.
We combined this with the temporal filtering given by the closed loop transfer function.
In Fig. \ref{fig:cumulative_RMS_jitter_speed} we reported the cumulative RMS of the jitter speed and its forecast error.

We can see in this figure that the error after the forecast is about half of the input, but it is concentrated at the high frequencies where the closed loop is able to reduce it effectively.
Note that the forecast is accurate on low frequencies and it introduces errors above 10Hz.
In particular, it magnifies disturbances at high frequencies like measurement noise, non-linearity due to elongated/truncated spots and possible vibrations in the launcher, because it comprises two derivatives in cascade, one to compute speed from position and one to compute acceleration from speed (see Eq.~\ref{eq:v0} and \ref{eq:a0}).
It is interesting to note that we could reduce the closed loop transfer function bandwidth, acting on the gain of the loop, to mitigate the distortion induced aberrations propagation but, unfortunately, this will impact on the ability of MORFEO to correct turbulence aberrations.
For the median atmospheric profile\cite{2013aoel.confE..89S} we expect a reduction of a factor 0.21 meaning that we should be able to mitigate the distortion induced aberration to 78nm from the original 370nm.
Note that actually the tomographic reconstructor is applied before the temporal filtering, but since they are linear operators we reverse their order in this analytical estimation (the propagation in the tomographic reconstructor is presented in Sec.\ref{sec:prop}).
It is worth noticing that the mitigation is mostly effective after the tomographic propagation when the temporal filter is applied, this means that the pseudo-open loop will be characterized by a strong disturbance. Hence all the tasks that use this data, like PSF reconstruction and atmospheric profile estimation, vibration peaks estimation will require additional data processing to deal with the residual distortion induced aberrations.

\section{Propagation of distortion induced aberrations in MORFEO}
\label{sec:prop}

The residual presented in the previous section is the one seen by a single WFS and, in MORFEO, it will propagate through the tomographic reconstruction and projection matrices.
Please note the propagation through the noise transfer function has already been considered in the previous section.
We compute the propagation coefficient (from WFS error to on-axis error) in the MORFEO reconstruction and projection matrices for third order modes only, that are the dominant term of the distortion induced aberrations and we found 0.805 and 0.795 for M4 and for the post focal DM (we are considering the single post focal DM case) respectively. These coefficients were determined by computing modal reconstruction and projection of 1000 random realizations of coma and trefoils with unitary RMS value.
Then, combining the DIA presented in Sec.~\ref{sec:mitigation} and the propagation coefficients we have 62nm per mirror that is a total of approximately 90nm.

\section{Performance estimation}
\label{sec:performance}

\begin{figure}[h]
    \centering
    \includegraphics[width=0.65\linewidth]{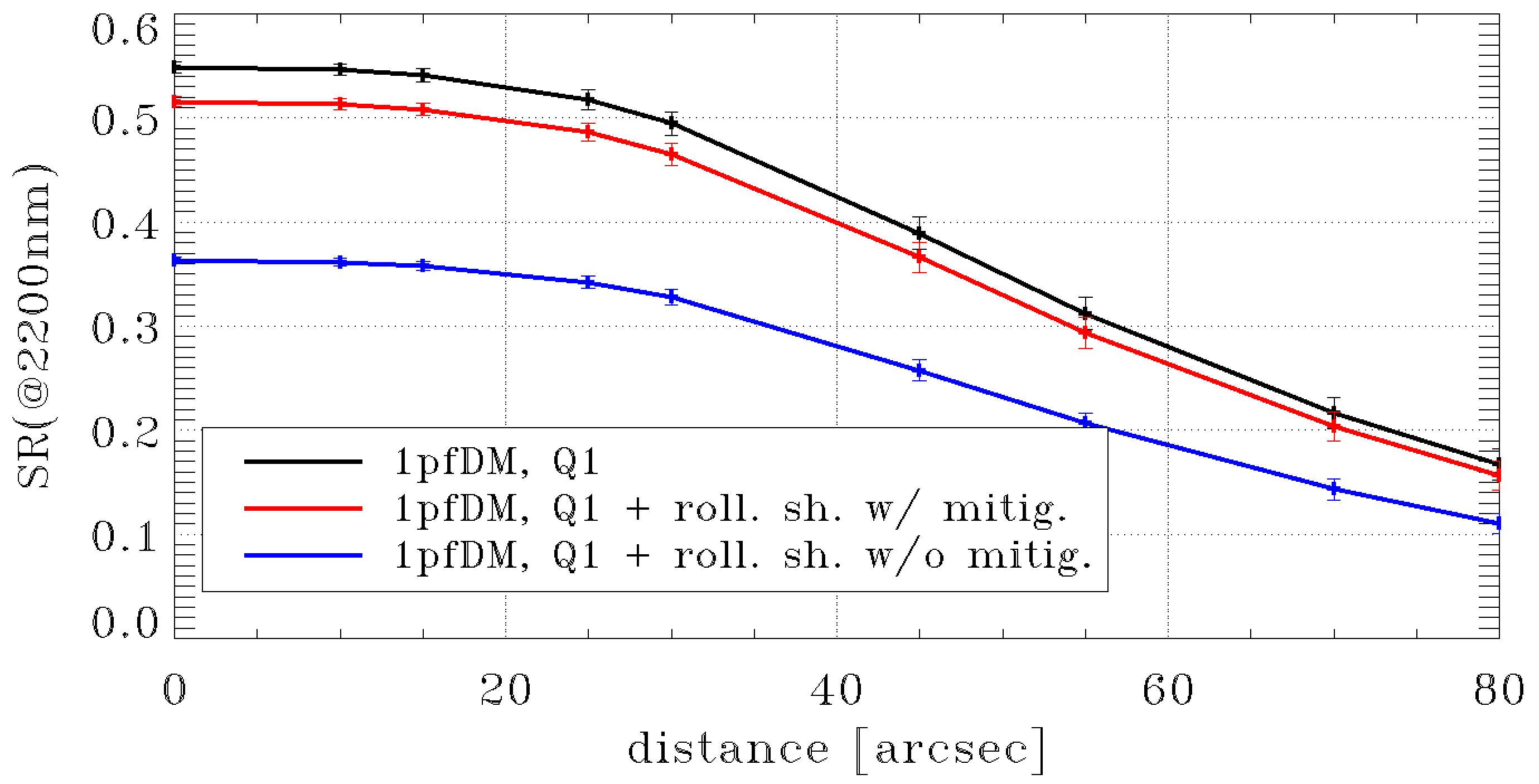}
    \caption{K band SR in the FoV with a rolling shutter detector (with and without mitigation) with respect to the baseline case with a global shutter camera and a single post focal DM.}
    \label{fig:Q1_perf}
\end{figure}

\begin{figure}[h]
    \centering
    \subfigure[Large MICADO FoV (53 $\times$ 53 arcsec).\label{fig:median_sky_cov_large}]
    {\includegraphics[width=0.75\columnwidth]{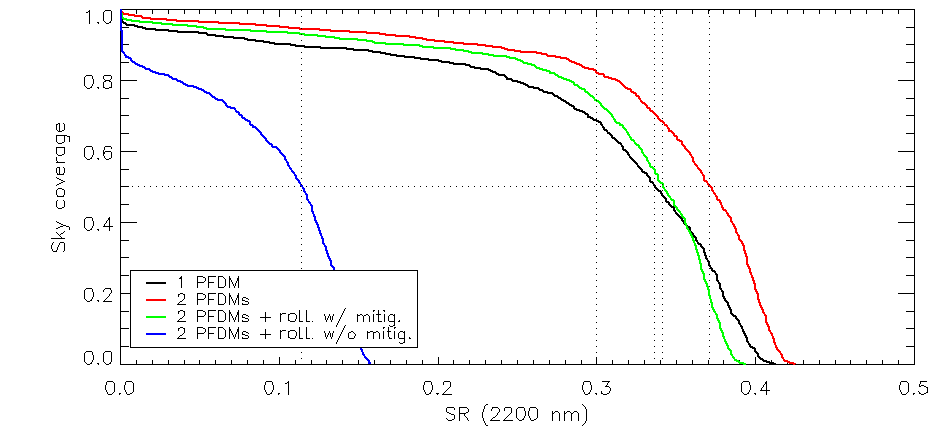}}
    \subfigure[Small MICADO FoV (20 $\times$ 20 arcsec).\label{fig:median_sky_cov_small}]
    {\includegraphics[width=0.75\columnwidth]{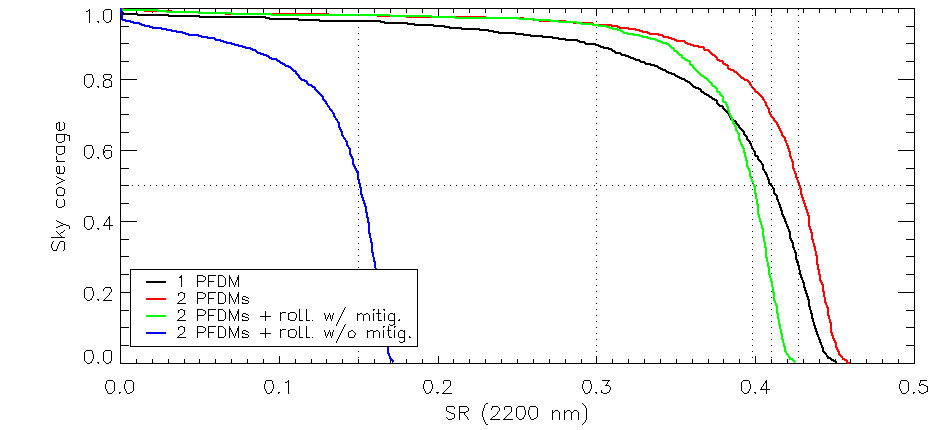}}
    \caption{Sky coverage for MICADO for the median profile\cite{2013aoel.confE..89S}, comparing the impact of a rolling shutter detector (with and without mitigation) with respect to the impact of the 2nd post focal DM. The use of a global shutter detector cameras is considered in black and red line\label{fig:median_sky_cov}}
    
\end{figure}

As reported in the previous section the impact on the LGS loop due to rolling shutter read-out is about 90nm, but it could be as high as 350m in the worst case scenario (that is without mitigation).
An additional error of 90nm means that K band SR scales of 0.94, H band SR of 0.89 and J band SR of 0.81 respectively.
Then, we consider also 40nm of additional error due to other differences of the LISA and SONY detectors: number of pixels (800 vs 1100) and quantum efficiency (0.95 vs 0.7).
The number of pixels allow for a larger number of sub-apertures with the same sub-aperture FoV, but this gives an improvement lower than the one expected with a single conjugate system due to the super-resolution\cite{2022JATIS...8b1514F}, while quantum efficiency is better for the LISA camera.
So the final difference becomes about 100nm.
The effect of this 100nm can be seen in Fig. \ref{fig:Q1_perf}, where we report the expected SR in the best atmospheric conditions (seeing 0.43arcsec at zenith).
We evaluate the impact of such additional error on the sky coverage considering also the reduced Signal-to-Noise Ratio (SNR) on the LO WFS given by the lower H band SR on the technical FoV\footnote{MORFEO NGS works in H band, see Ref. \citeonline{2018SPIE10703E..4DB}.}.
Note that this is a simplified (see Ref. \citeonline{2022JATIS...8b1509P} for a description of the sky coverage estimation method) and optimistic approach, because the correlation between correction in the technical FoV and LO correction is not limited to the SNR only.
We decided to focus on comparing the impact of a rolling shutter detector with respect to the impact of the 2nd post focal DM.
We can see in Fig. \ref{fig:median_sky_cov} that for half of the pointings (40\% for the large MICADO FoV, 70\% for the small MICADO FoV) where we expect the best performance MORFEO with a single DM performs better than MORFEO with the 2nd post focal DM and with a rolling shutter detector with the mitigation.
This means that we lose the advantage of the second post focal DM in particular for Galactic Astronomy and Resolved Stellar Populations.
The 2nd post focal DM holds its advantage at high sky coverages so, in particular, for Extragalactic Astronomy.
Moreover, we verified that the 2nd post focal DM is not able to compensate for the impact of a rolling shutter detector in the best atmospheric conditions.
Hence, while it is possible to fulfill the requirements, because we have a K band SR$>50\%$ for the best atmospheric conditions and $>30\%$ for a sky coverage of 50\%, the impact of a rolling shutter detector is greater than the one given by the 2nd post focal DM in about half of the observations, in particular the ones where we expect the best performance from MORFEO.


\section{Conclusion}
\label{sec:conclusion}

We have analyzed the effect of the image distortion in the LGS WFS of MORFEO equipped with a detector using a rolling shutter read-out scheme.
These rolling shutter CMOS detectors are particularly attractive for MORFEO because of their large format, low-noise and low-latency features.
A typical residual LGS jitter of 100-200 mas, although negligible in term of tilt signal for an ELT LGS WFS, corresponds to a large residual aberration of a few $\mu m$ RMS and propagates rolling shutter induced aberration of more than 300nm RMS on modes of the third radial order and above.
Mitigation combined with the closed loop filtering is able to reduce this distortion induced aberration by a factor 5, but after the tomographic reconstruction and projection we have still 90nm of additional error.
We evaluate the performance of MORFEO with such additional error and we verify that while requirements are still met the advantage given by a second post focal DM is almost cancelled.
Therefore MORFEO consortium asked ESO to adopt as LGS WFS detector a camera based on SONY IMX 425 chip. 

\subsection* {Acknowledgments}

We acknowledge Fernando Gago for sharing with us the mitigation strategy presented in this work and Sylvain Oberti for sharing with us GALACSI NFM jitter data.\\
This work has been partially funded by ADONI – the ADaptive Optics National laboratory of Italy.


\bibliography{report}   
\bibliographystyle{spiejour}   

\end{document}